\documentclass[11pt,twoside]{article}

\usepackage{asp2006}
\usepackage{epsf}
\usepackage{psfig}
\usepackage{lscape}
\usepackage{graphicx}
\usepackage{color}
\begin{document}
\title{Dust in Supernovae; Formation and Evolution}   

\author{Takashi Kozasa$^1$, Takaya Nozawa$^2$, Nozomu Tominaga$^3$, 
Hideyuki Umeda$^4$, Keiichi Maeda$^2$ \& Ken'ichi Nomoto$^{2,4}$ } 
\affil{${^1}$Department of Cosmosciences, Graduate School of Science, Hokkaido
University, Sapporo 060--0810, Japan\\
$^{2}$Institute for the Physics and Mathematics of the Universe, University of
Tokyo, Kashiwa 277-8568, Japan \\    
$^{3}$Optical and Infrared Astronomy Division, National Astronomical 
Observatory of Japan, Mitaka 181-8588, Japan \\
$^{4}$Department of Astronomy, School of Science, University of Tokyo, 
Tokyo 113-0033, Japan}

\begin{abstract} 
Core--collapsed supernovae (CCSNe) have 
been considered to be one of 
sources of dust in the universe.  What kind and how much mass of dust 
are formed in the  ejecta and are injected into the interstellar medium
 (ISM) depend on the 
type of CCSNe, through the difference in 
the thickness (mass) of outer envelope.
In this review, after summarizing the existing 
results of observations on dust formation in 
CCSNe,  we investigate formation of dust  
in the ejecta and its evolution in the 
supernova remnants (SNRs) of  
Type II--P and Type IIb SNe. Then, 
the time evolution of thermal emission 
from dust in the SNR of Type IIb SN is demonstrated and compared 
with the observation  of Cas A.
We find that the total dust mass formed in the
 ejecta does not so much depend on the type; $\sim 0.3-0.7 M_{\odot}$ in 
Type II--P SNe and $\sim 0.13 M_{\odot}$ in Type IIb SN. 
However the size of dust sensitively depends on the type, 
being affected by the difference in the gas density in the ejecta: 
the dust mass is dominated by grains with radii  
larger than 0.03 $\mu$m in 
Type II-P, and less than 0.006 $\mu$m in Type IIb, which 
decides the fate of dust in the SNR. The surviving dust 
mass is $\sim 0.04-0.2 M_{\odot}$ in the SNRs of 
Type II--P SNe for
the ambient hydrogen density of $n_{\rm H}=10.0-1.0$ cm$^{-3}$, 
 while 
almost all dust grains are destroyed in the SNR of Type IIb.
The spectral energy distribution (SED) of thermal emission from 
dust in SNR well reflects the evolution of dust grains in SNR through
erosion by sputtering and stochastic heating. 
The observed SED of Cas A SNR is 
reasonably reproduced by the model of dust formation and evolution for 
Type IIb SN. 
\end{abstract}



\section{Introduction}

Dust is one of important ingredients in space to control the physical 
and chemical conditions in interstellar medium (ISM) and the formation
process of stars in molecular clouds via the interactions with radiation 
and gas as well as the  
formation of H$_2$ molecule on the surface. In astrophysical
environments, dust grains form in a cooling gas outflowing 
from star to interstellar space such as 
in the stellar winds from asymptotic giant branch (AGB) 
stars and in the
ejecta of novae and supernovae (SNe) where gas density is high enough to 
proceed the formation and growth of seed nuclei through collisions of 
relevant gas species. 

Formation of dust in the ejecta of SNe  
had been invoked to explain the origin of interstellar dust
(Cernuschi, Marsicano, \& Codina 1967), 
and the microscopic analysis of 
individual grains in meteorites has identified the dust grains 
formed in SNe 
from their
isotopic compositions (see Clayton \& Nittler 2004 for a review). 
Dust formation in the ejecta of SNe had been directly
observed for the first time in SN 1987A (see
McCray 1993 for a review), after then dust formation in 
the ejecta was reported
in a handful of core--collapsed SNe (CCSNe).  
Based on the observations of nearby SNe, 
the mass of dust 
formed in the ejecta has been considered to be less than $10^{-3}
M_{\odot}$ per SN, which is more than two to three orders of magnitude smaller 
than the dust mass predicted by theoretical
investigations 
(Kozasa, Hasegawa, \& Nomoto 1989, 1991; Todini \& Ferrara 2001; Nozawa
et al. 2003). Furthermore the observed dust mass is too small to explain the
amount of dust observed in the host galaxies of quasars at redshift of  $z>5$ 
(e.g. Bertoldi et al. 2003) 
where CCSNe are considered to be a  major source of dust because 
the cosmic age is too young for AGB stars to supply the dust. 
So how much mass of dust forms in the ejecta has been still 
controversial. 
 
In the ejecta of SNe, dust
formation proceeds within the 
He--core where condensible elements are
more abundant and the gas density is higher enough  due 
to the smaller expansion velocity than in the outer envelope.  
How much mass and what kind 
of dust
are injected from a SN into ISM depend not only on  the
formation in the ejecta but also on the destruction 
in the supernova remnant (SNR) where dust grains 
are injected into and eroded by sputtering in the hot gas   
swept up by the reverse and forward shocks produced 
by the interaction of ejecta 
with interstellar/circumstellar medium. 
Both processes depend on the type of 
SNe  through the thickness (mass) of outer envelope: With the
same kinetic energy of explosion,  the
thicker outer envelope makes the expansion velocity 
slower and results in the gas density within 
He--core being high enough to form 
large--sized dust. Furthermore,  
the lower gas density in the shocked region 
caused by the delayed arrival of the reverse shock at the  
dust forming region decreases the 
erosion rate of dust by sputtering. 

In this review we 
show how the formation and destruction of dust in SNe   
depend on the type, comparing the results of calculations
for Type II--P and Type IIb SNe; Type II--P SN 
is a typical type among CCSNe, and the well studied Cas A 
was  
recently 
identified as
Type IIb SN (Krause et al. 2008). Then, we demonstrate the time
evolution of thermal 
emission from dust residing in SNR of Type IIb SN, 
based on the results of calculations of dust formation and destruction. 
In \S ~2, we summarize the observations of dust formation in SNe 
after briefly introducing the classification of SNe. We present the 
results of calculations of dust formation in the ejecta of 
Type II--P and IIb SNe 
in \S ~3. Focusing on Type IIb SN, in \S ~4 we calculate the evolution of
dust in the SNRs and the time evolution of thermal emission
from dust,  and compare with 
the observation of
Cas A. Summary and concluding remarks are presented in \S ~5.

\section{Observations of dust formation in supernovae}

In this section we summarize the observations of dust formation in
SNe. First we briefly introduce the classification of SNe   
for your information, and then review  
the observations of dust forming SNe. Massive 
stars whose main sequence mass is larger than 8 $M_{\odot}$
end up their lives as CCSNe, 
but the details depend on the initial metallicity 
(Heger et al. 2003 for details). Here 
we shall confine to SNe evolved from stars with solar 
metallicity.   

\subsection{Classification of supernovae}

Supernova is the most energetic explosion in the
universe. Physically the supernovae are classified by the
explosion mechanism; Core--collapsed supernovae  
which are triggered by the gravitational collapse of 
the central core, 
and thermonuclear supernovae in which the explosion 
energy is supplied by explosive thermonuclear burning
(e.g. Hillebrandt \& Niemeyer 2000). 

Observationally SNe are classified by the 
spectral features at the early phase and the behavior of 
the light curve (e.g. Wheeler \& Harkness 1990; 
Filippenko 1997; Leibundgut 2008). 
According to the presence of the H feature, 
SNe are divided into Type I
and Type II; 
Type I without H, and
Type II with H. Type I SNe are classified by the strength of 
the Si feature
(Si II absorption at 6150 \AA); Type Ia SNe 
showing the strong Si absorption feature are thermonuclear SNe 
whose progenitors are white dwarfs in binary systems.  
H--deficient Type I SNe with weak Si feature are 
subdivided by the
content of He. He--rich is Type Ib and He--poor is Ic. The progenitors 
are Wolf--Rayet stars; H--envelope, and both H-- and He--envelopes 
are removed before explosion for Type Ib SN and Type Ic SN, respectively, 
by stellar wind and/or interaction in binary
system.  

Type II SNe, whose progenitors are red supergiants, are 
 divided by the content of H; H--poor are 
IIb SNe in which almost all of the  
H--envelope is removed during the evolution. H--rich Type II 
SNe are classified by the behavior of the  
light curves; the light curve 
linearly decays in Type II--L SNe, and have 
a plateau in Type II--P SNe. 
Type II SNe showing narrow emission lines originating 
in circumstellar medium are labelled as IIn. Type II, Ib and 
Ic SNe are core--collapsed supernovae.

\subsection{Observations of dust formation in core--collapsed supernovae}
 
Dust formation in the ejecta has been observed so far in the ejecta of 
CCSNe except for Type Ic SNe. No evidence of dust
formation is reported in the ejecta of Type Ia SNe. 
Table 1 summarizes the
observations of dust formation in the ejecta of CCSNe.
Formation of dust in the ejecta had been confirmed
observationally for the first time in SN 1987A which is a peculiar Type
II SN because the progenitor is not a red supergiant but a
blue supergiant (West el al. 1987). 

The evidences of dust 
formation in the ejecta of SN 1987A (see McCray 1993; Wooden 1997; 
the references in Table 1) are the 
following; 
(1) 
the enhancement of infrared flux 
at $\sim$ 10 $\mu$m starting from day 450 followed by 
the decline of U to M bolometric 
 luminosity from $\sim$ day 500, and (2) the appearance of 
blue--shifted profiles of [Mg I], [O I], [C I] lines starting 
from $\sim$ day 530. 
The behavior of bolometric luminosity 
obtained by combining  
the observed U to M, infrared (thermal
emission from dust), X-ray 
and $\gamma$--ray luminosities satisfies the energy budget expected from 
the decay of radioactive element $^{56}$Co. 
The appearance of blue--shifted lines was interpreted as the
result of the attenuation of emission from the far (receding) side of the 
ejecta caused by dust grains formed in
clumps within the ejecta, and in addition the intensities of 
[Mg I], [O I] and [Si I] lines concurrently decreased (Lucy et
al. 1991). It should be noted
 that, in prelude to the dust formation, formation of CO and SiO molecules
 was observed as early as day 112 and 165, respectively. 
The emission of SiO was no longer 
recognized at day 517 around which dust formation 
was confirmed by the observation of 
blue--shifted lines (Roche et al. 1991). 
The dust mass estimated from the observation was at least 
$1 \times 10^{-4} M_{\odot}$ at day 775, and no information on the dust
composition was available since the thermal emission was well fitted by
a grey body (Wooden et al. 1993).

Taking the appearances of the blue--shifted lines, infrared excess and/or 
decline of optical luminosity as 
the diagnostics, so far 
dust formation in the ejecta has been reported in four SNe excluding SN
1987A (see the references in Table 1); 
three are Type II--P SNe  
which is the most common type  
among CCSNe (e.g. Smartt et al. 2009), and one is Type Ib. 
The evidence of 
dust formation in SN 2005af is based only on    
the mid--infrared (MIR) excess observed by Spitzer, but the detection of CO
and SiO molecules supports the formation since CO and/or SiO 
molecules are detected prior to dust formation in 
Type II--P SNe. The molecules 
could play an important role in
formation of dust because CO acts as a coolant in 
the ejecta (e.g. Liu \& Dalgarno 1995), and 
SiO is considered to be a precursor of 
dust grains (Kozasa et al. 1989).  The onset of dust formation
ranges from $\sim 300$ to $\sim 600$ days after 
the explosion in Type II--P SNe. In SN 1990I,
which is Type Ib, the onset seems to be earlier, as can be expected,  
because the escape of 
$\gamma$-rays with 
less effective deposition causes the gas in 
the ejecta to cool down faster without an  
H--envelope.
The dust mass derived from the observations is less than $10^{-3}
M_{\odot}$, although the observation of SN 1999em  
suggested that
dust much more than $10^{-4} M_{\odot}$ was produced in the ejecta 
(Elmhamdi et al. 2003). 
 
Recently formation of dust in cool dense shells (CDSs) 
generated by the
interaction of ejecta with a  
circumstellar medium (CSM) 
has been reported in 
three SNe summarized in Table 2 (see the references in Table 2);  
two are Type IIn and one is  Type Ib,
where the presence of dense circumstellar envelopes is  
confirmed by the 
X--ray observations (e.g. Smith et al. 2008a). 
The evidences come from the NIR/MIR excess and the 
concurrent appearance of the blueshift of the narrow lines in 
CSM and/or the  
intermediate--width components originating in shocked gas. 
Onset of dust formation depends on when the ejecta
encounters with dense CSM. The early formation of dust at day $\sim 50$ in
SN 2006jc is consistent with the LBV--like outburst $\sim 2$ yr  
prior to the explosion. 
The dust mass estimates  
are $>2 \times 10^{-3} M_{\odot}$ in SN 1998S, and $3
\times 10^{-4}$ by day 230 in SN 2006jc, assuming carbon
dust. It should be noted that in addition to the formation of dust in CDSs, 
we cannot deny the possibility that dust also forms  
in the expanding
ejecta; see Smith et al. (2008b) for SN 2005ip; 
Sakon et al. (2009), Tominaga et al. (2008), and Nozawa et al. (2008)
for SN 2006jc. Although the frequency of Type IIn SNe is rare among CCSNe
(Gal--Yam et al. 2007; Smartt et al. 2009), these observations on dust 
formation in CDS  
provide a new window to investigate the formation process of 
dust in astrophysical environments.
 
Nowadays, based on the observations, the  
mass of dust formed in the ejecta
has been claimed to be less than $10^{-3} M_{\odot}$ and to be 
too small to contribute to the inventory of dust in our Galaxy. 
However, we should keep in mind that the
conclusion is not necessarily 
definite, based on the
limited number of the observations and the assumption that 
thermal radiation from dust is optically 

\begin{landscape}
\begin{table}[!ht]
\caption{Formation of dust in the ejecta. References; 
1) Lucy et al. (1991), 2) Meikle et al. (1993),  3) Roche et
 al. (1989), 4) Whitelock et al. (1989), 5) Suntzeff \& Bouchet (1990), 
6) Lucy et al. (1989), 7) Catchpole \& Glass (1987), 8) Aitken et
 al. (1988), 9) Ercolano et
 al. (2007), 10) Elmhamdi et al. (2004),
 11) Elmhamdi et al. (2003), 12) Spyromilio et al. (2001), 
13) Sugerman et al. (2006), 14) Hendry et al. (2005), 15) Meikle et
 al. (2007), 16) Kotak (2008), 17) Kotak et al. (2006).
 Note; The optical depth of the dust core 
derived from [O I]
 line profile $\tau_{\rm d} \gg 10$ at day 510 in SN 1999em  
 (Elmhamdi. et al. 2003), 
which is much larger than $\tau_{\rm d} \sim 1$ in SN 1987A (Lucy et al.
 1991).}
\vspace{-0.5cm}
\smallskip
\begin{center}
{\smallskip
\begin{tabular}{llllllll}
\tableline
\noalign{\smallskip}
name & Type & onset of dust  & IR excess & decline of   & blue--shifted &
 molecules & $M_{\rm d}$ \\
\null & \null & formation $t_{\rm c}$[day] & (NIR/MIR) & opt. lum. & lines & \null &
 [$M_\odot$] \\ 
\noalign{\smallskip}
\tableline
\noalign{\smallskip}
SN 1987A & II--pec. & $\sim$ 350--530 $^{1), 2)}$ & 
MIR $^{3)}$ & yes $^{4), 5)}$ & [Mg I],[O I] etc. $^{6)}$  & CO $^{7)}$, SiO 
 $^{8)}$ &
 $7.5\times 10^{-4}$ \ $^{9)}$ \\ 
SN 1990I & Ib & $\sim 250$ \ $^{10)}$ & -- & yes $^{10)}$ & [O I], [Ca II]
 $^{10)}$&
 -- & -- \\ 
SN 1999em & II--P & 465$<t_{\rm c}<$510 $^{11)}$ & -- & yes $^{11)}$  
 & [O I],[Mg I],etc. $^{11)}$ & CO $^{12)}$ & $\ge 10^{-4}$ \ $^{11)}$\\ 
SN 2003gd &II--P & 250$<t_{\rm c}<$493 $^{13)}$& MIR $^{13)}$& yes
 $^{13)}$ 
& [O I],
 H$_{\alpha}$ $^{13), 14)}$& CO,SiO ? $^{15)}$&
 $4\times 10^{-5}$ \ $^{15)}$ \\ 
SN 2005af & II--P & 214$<t_{\rm c}<$571 $^{16)}$ 
& MIR $^{16)}$& -- & -- & CO,SiO $^{17)}$ 
& $\sim 4\times 10^{-4}$ \ $^{16)}$\\ 
\noalign{\smallskip}
\tableline
\end{tabular}
}
\end{center}
\end{table}
\vspace{0.0cm}
\begin{table}[!hb]
\vspace{-1.0cm}
\caption{Formation of dust in a cool dense shell produced by the
 interaction with dense circumstellar medium. References;  
1) Pozzo et al. (2004), 2) Gerardy et al. (2000), 
3) Smith et al. (2008b),
4) Fox et al. (2009), 5) Smith et al. (2008a), 6)Mattila et al. (2008).
Note: For SN 2005ip, Fox et al. (2009) suggested the early formation of
 dust at day $\sim 50$, but the spectroscopic observation (Smith et al.2008b) 
suggested the later formation. It should be noted that
Smith et al. (2008b) have suggested formation of dust in the
 ejecta at earlier time.}
\smallskip
\begin{center}
{\small
\begin{tabular}{llllllll}
\tableline
\noalign{\smallskip}
name & Type & onset of dust & IR excess & decline of  & blue--shifted &
 molecules & $M_{\rm d}$ \\
\null & \null & formation $t_{\rm c}$ [day] & (NIR/MIR) & optical lum. & lines & \null &
 [$M_\odot$] \\ 
\noalign{\smallskip}
\tableline
\noalign{\smallskip}
SN 1998S & IIn \hspace{1cm} & $\sim250$ $^{1)}$ \hspace{0.8cm} 
& NIR $^{1)}$ & --  &
 H$_{\alpha}$, He I $^{1)}\hspace{1.0cm}$ & CO $^{2)}$\hspace{2.0cm} &
 $>2 \times 10^{-3}$ $^{1)}$ \\ 
SN 2005ip & IIn & $\sim 730$ $^{3)}$  & NIR $^{3,4)}$ & -- & He I
 $^{3)}$ & -- & -- \\ 
SN 2006jc & Ib & $\sim 50$ $^{5,6)}$& NIR--MIR $^{5,6)}$& yes &  
He I $^{5,6)}$& -- & $3 \times
 10^{-4}$ \ $^{6)}$ \hspace{2.0cm}\\ 
\noalign{\smallskip}
\tableline
\end{tabular}
}
\end{center}
\end{table}
\end{landscape}

\noindent
thin in some cases;  
as pointed out by Meikle et al. (2007), the dust may reside 
in optically thick clumps in the ejecta.
Also, the NIR--MIR observations could miss the cool dust, since 
dust grains might cool down quickly after the formation
as is demonstrated by Nozawa et al. (2008). Therefore, 
the monitorings of various   
types of SNe   
after the explosions, covering the
temporal and wavelength ranges as wide as possible, are necessary to 
reveal the mass of dust formed in the ejecta and 
its dependence on
the type of SNe. In addition the 
sophisticated radiative transfer
calculations such as those carried out by Sugerman et al. (2006) and
Ercolano et al. (2007) are inevitable   
to derive the dust mass from the observations relevant to dust 
formation, taking into account the more realistic spatial distribution and
chemical composition of dust in the ejecta.
 
\section{Formation of dust in the ejecta}

In this section we investigate how the formation process of dust 
in the ejecta depends on the type of SNe,  
focusing on Type II--P and Type IIb SNe whose progenitors are 
evolved from massive stars with solar metallicity. 
The calculation of dust formation is
based on the theory of nucleation and growth developed by Nozawa et al
 (2003), taking into account chemical reaction at 
the condensation by
considering that the key species, defined as a gas species with 
the least 
collisional frequency among the reactants, controls the kinetics of
nucleation and grain growth (Kozasa \&
Hasegawa 1987, Hasegawa \& Kozasa 1988) \footnote{There was a typo in the
nucleation rate $J$ in Kozasa \& Hasegawa (1987) and Nozawa et al.
(2003); replace $c_1$ with $c_1^2$, where 
$c_1$ is the number density of key species.}. Given the time evolution    
of gas density and temperature together with 
the elemental composition, 
we can determine when, where, what kind, what size, and how much mass of  
dust condenses in the ejecta. 

\subsection{Models of supernovae and dust formation calculation}

Formation process of dust grains in the ejecta of SNe
is controlled by the time evolution of 
gas density and temperature as well as the elemental composition in the
ejecta as demonstrated  
by Kozasa et al. (1989). Thus, we apply the 
hydrodynamic model of a  
SN explosion for the time evolution of the  
gas density  
and the nucleosynthesis calculation for the elemental composition in 
the ejecta.
The time evolution of the gas
temperature is calculated by solving the radiative transfer equation
taking into account the energy deposition from radioactive elements.
Table 3 summarizes the models of SNe  
used for the calculations.

\begin{table}[!ht]
\caption{Models of SNe; $M_{\rm pr}$ is the progenitor mass,
$E_{\rm exp}$ the kinetic energy of the explosion,
 $M(^{56}{\rm Ni})$ the mass of $^{56}$Ni in the ejecta, $
M_{\rm eje}$ the mass of the ejecta, 
$M_{\rm H-env}$ the mass of the hydrogen envelope, 
$M_{\rm He-core}$ the
 mass of the He--core, and $M_{\rm matal}$ 
the mass of metal inside the He--core. Model A is 
taken from Nomoto et al.
 (2006), and model B from Umeda and Nomoto (2002).}
\smallskip
\begin{center}
{\small
\begin{tabular}{ccccccccc}
\tableline
\noalign{\smallskip}
model & Type & $M_{\rm pr}$ & $E_{\rm exp}$ &
 $M(^{56}{\rm Ni})$ & $M_{\rm eje}$ & $M_{\rm H-env}$ & $M_{\rm
 He-core}$ & $M_{\rm metal}$\\ 
\null & \null & [$M_{\odot}$] & [erg] & [$M_{\odot}$] &  [$M_{\odot}$] 
& [$M_{\odot}$] & [$M_{\odot}$] & [$M_{\odot}$]\\
\tableline
\noalign{\smallskip}
A & II--P & 15.0 & $10^{51}$ & 0.134 & 12.2 & 10.4 & 1.80 & 0.482\\
B & II--P & 20.0 & $10^{51}$ & 0.07 & 15.7 & 10.9 & 4.77 & 3.35\\
C & IIb  & 18.0 & $10^{51}$ & 0.07 & 2.94 & 0.08 & 2.86 & 1.297\\  
\noalign{\smallskip}
\tableline
\end{tabular} }
\end{center}
\end{table}

\begin{figure}[!th]
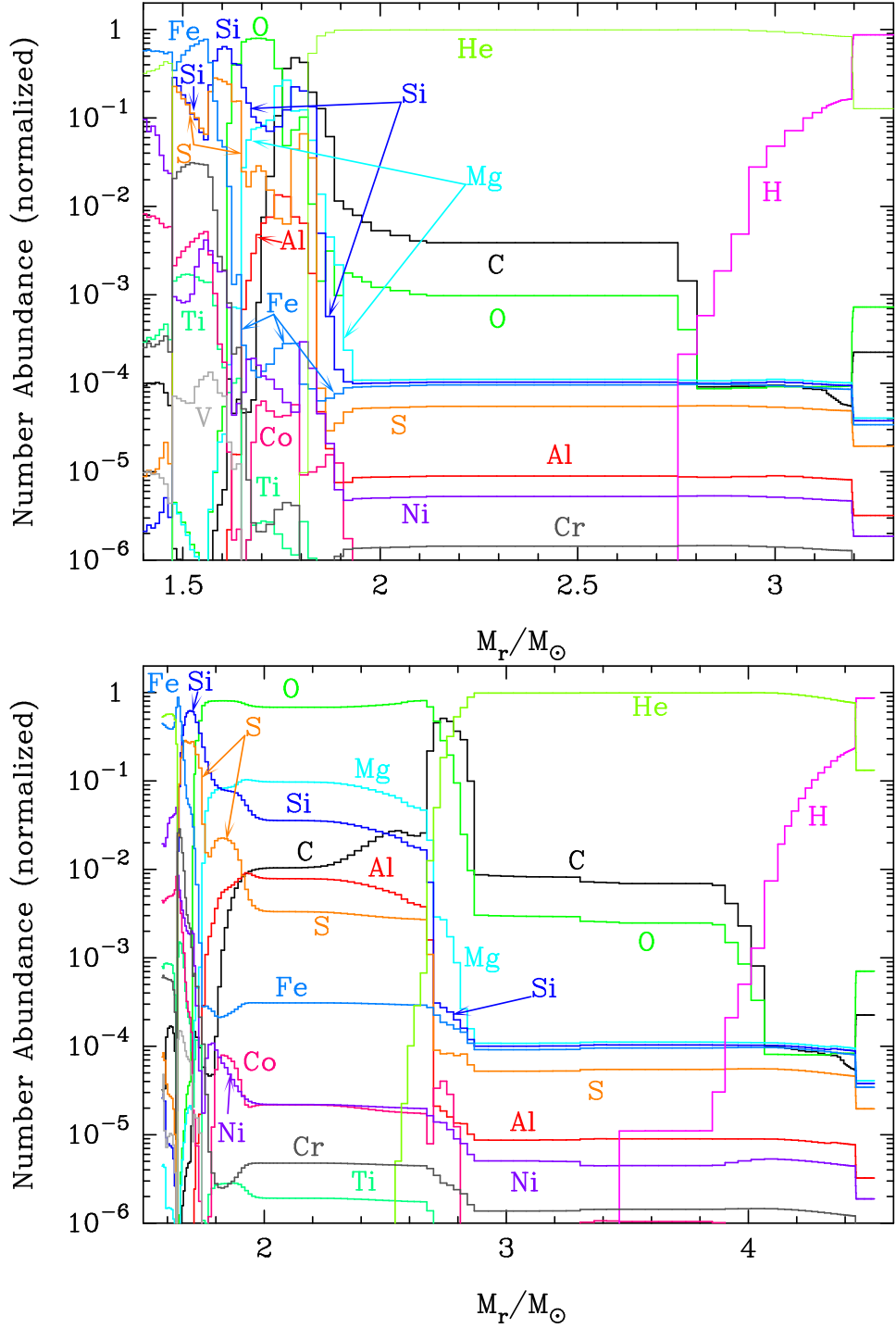

\begin{center}
  \includegraphics[scale=.3, angle=270, totalheight=270pt, clip=true]
{figcolor/abun.15m.cl.ps}
  \includegraphics[scale=.3, angle=270, totalheight=270pt, clip=true]
{figcolor/abun.93j.cl.ps}
\caption{Elemental composition within He--core: Upper panel for Type 
 II--P SN (model A) and lower panel for Type IIb SN (model C). 
Note that the position
 in the ejecta is indicated by the mass coordinate that is defined as the
 enclosed mass from the center.}
\vspace{-1.0cm}
\end{center}
\end{figure}

In the calculations we consider that dust forms  
within the He--core
because the large expansion velocity in the outer envelope causes  
the density of condensible elements to be too low 
to form dust grains. For the elemental
composition, we do not consider any mixing since the mixing driven by 
the Rayleigh--Taylor instability in the ejecta at the
explosion is not microscopic but 
macroscopic as have been revealed by the observation of Cas A 
(Douvion et al. 1999; Ennis et al. 2006).
We assume that formation of CO and SiO molecules is complete; all carbon
(silicon) atoms are locked into CO (SiO) molecules in  the locations   
where C/O (Si/O) is less than 1. This implies that C--bearing 
dust condenses only in the region of C/O$>1$; see 
Todini \& Ferrara (2001) and Bianchi \& Schneider
(2007) for the calculation of dust formation assuming the uniform
mixing of elements and considering formation processes of 
CO and SiO in the ejecta.

Figure 1 shows the elemental composition within the  
He--core for a Type II--P
SN (model A; upper panel) and for a  
Type IIb SN (model C; lower panel). 
From the view point of
dust formation, the  
He--core is  roughly divided into four layers; the outer 
C--rich layer, the O--rich layer, the 
Si--S--Fe layer 
and the Fe--S--Si layer.  
What kind of dust grain really condenses depends on the details of 
elemental composition. Thus, 
in the calculation we simultaneously solve the
equations of nucleation and grain growth for 19 possible condensates 
(see Nozawa et al. 2003 for details).

Figure 2 shows the time evolution of the  
gas density and temperature in the
ejecta; Left panel for Type II--P SN (model A) and right panel for 
Type IIb SN (model C). 
It should be noted that the gas density in the ejecta of Type IIb SN 
is more than three orders of magnitude smaller than that of type II--P SN,
because the expansion velocity in  
the ejecta is much higher in Type IIb SN without thick H--envelope. 
The  
time evolution of the gas temperature 
depends on the model through the 
mass of H--envelope and $M(^{56}{\rm Ni})$ in the ejecta.

\begin{figure}[!h]
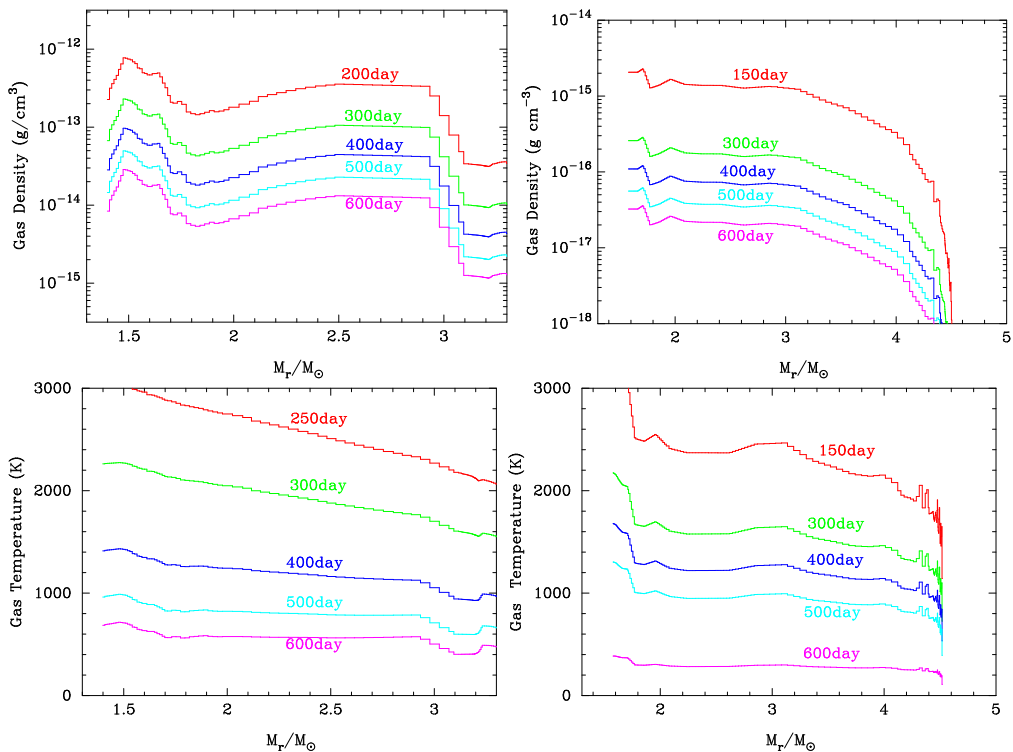

  \includegraphics[scale=.3, angle=270, totalheight=140pt, clip=true]
{figcolor/densi.15m.cl.ps} 
  \includegraphics[scale=.3, angle=270, totalheight=140pt, clip=true]
{figcolor/densi.93j.cl.ps}
  \includegraphics[scale=.3, angle=270, totalheight=140pt, clip=true]
{figcolor/temp.15m.cl.ps}
  \includegraphics[scale=.3, angle=270, totalheight=140pt, clip=true]
{figcolor/temp.93j.cl.ps}
\caption{Time evolution of gas density and temperature in the ejecta;
 Left panel for Type II--P(model A) and right panel for Type IIb (model C)}
\vspace{-0.0cm}
\end{figure}
\begin{figure}[!h]
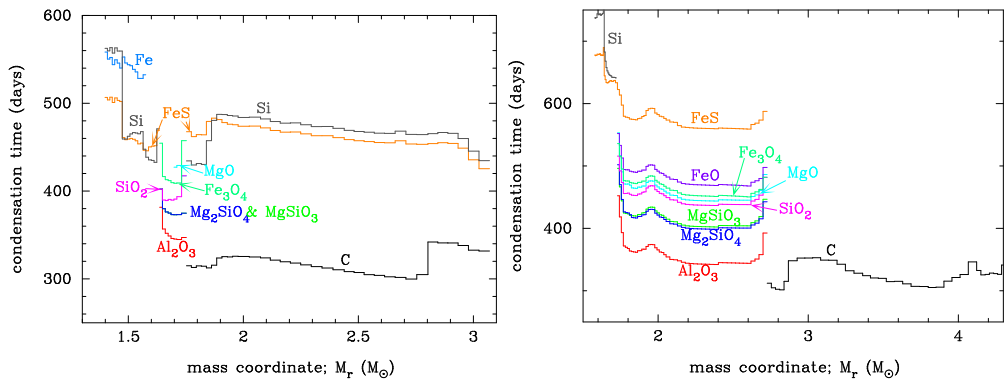

  \includegraphics[scale=.3, angle=270, totalheight=140pt, clip=true]
{figcolor/tc.15m.cl.ps} 
  \includegraphics[scale=.3, angle=270, totalheight=140pt, clip=true]
{figcolor/tc.93j.cl.ps}
\caption{Condensation times of dust species formed in the
 ejecta; Left for Type II--P SN (model A) and right for Type IIb SN (model C). 
The condensation time is
 defined as the time at which the nucleation rate reaches the  
 maximum. The condensation times of dust species 
in the C and O--rich layers are a little earlier in Type II--P than 
in Type IIb, except for FeS and
 Si in C--rich layer of Type II--P SN, and FeS in Type IIb }
\vspace{0.0cm}
\end{figure}

\subsection{Results of dust formation calculations}

Figure 3 shows the condensation time of each grain species formed in the 
ejecta; Left for Type II--P SN (model A) and right for Type IIb SN (model
C). As the gas in the ejecta cools down with time, C grains start to 
condense first in the C--rich layer around day 300 after the explosion. 
Afterwards, 
in the oxygen--rich layer, Al$_2$O$_3$, Mg$_2$SiO$_4$, MgSiO$_3$ and SiO$_2$ 
condense in this order from day 350 to day 450. In the Si--S--Fe rich layer, 
FeS and Si condense around day
450 in Type II--P SN and around day 630 in Type IIb SN. After then, 
in the innermost Fe--S--Si layer,
Si, FeS and Fe condense in Type II--P SN, but Si and FeS in Type IIb SN.  
The condensation sequence of MgO and Fe$_3$O$_4$ in the O--rich layer 
is different between Type II--P SN and Type IIb SN, depending on the detailed 
density structure and elemental composition in the ejecta. 
Also small amount of  FeS and Si condenses 
from 450 to 500 days in  the C--rich layer
of Type II--P SN, and so does FeS around 
day 580 in the O--rich layer of Type IIb SN. 

\begin{figure}[!h]
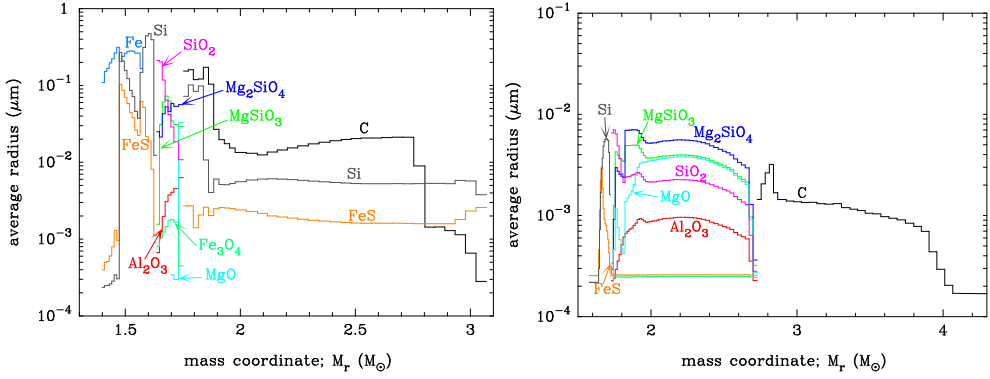

  \includegraphics[scale=.3, angle=270, totalheight=140pt, clip=true]
{figcolor/radi.15m.cl.ps}
  \includegraphics[scale=.3, angle=270, totalheight=140pt, clip=true]
{figcolor/radi.93j.cl.ps}
\caption{Average radii of dust species formed in the ejecta: Left for
 Type II--P SN (model A) and right for Type IIb SN (model C)}
\vspace{0.0cm}
\end{figure}
\begin{figure}[!h]
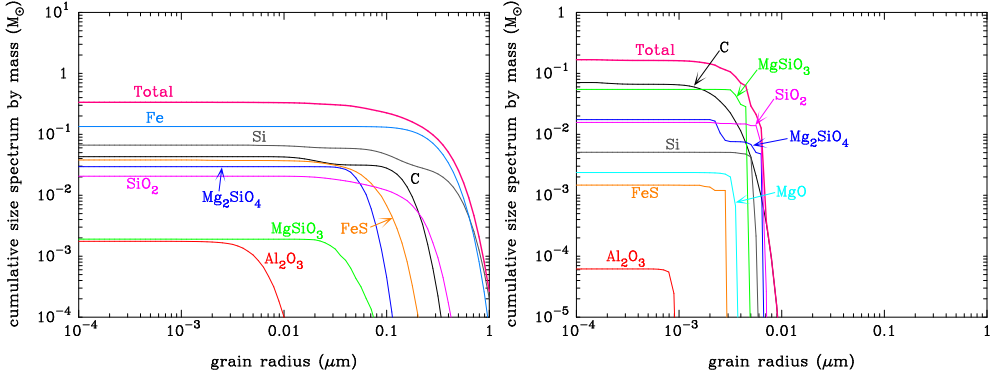

\includegraphics[scale=.3, angle=270, totalheight=140pt, clip=true]
{figcolor/size.15m.cl.ps}
  \includegraphics[scale=.3, angle=270, totalheight=140pt, clip=true]
{figcolor/size.93j.cl.ps}
\caption{Cumulative size distributions of dust species 
 by mass summed up over the
 ejecta. The mass is dominated by dust grains with radii larger than 
 0.03 $\mu$m
 for Type II--P SN (model A; left), but with radii less than 0.006 $\mu$m 
for Type IIb SN (model C; right)}
\vspace{0.0cm}
\end{figure}

Figure 4 shows the average radius of each dust species formed in the 
ejecta. In Type IIb SN (right panel), the average radii of all grain 
species are smaller 
than 0.01 $\mu$m due to the low gas density  in the ejecta. 
On the other hand, the average radii of dust grains are 
larger than 0.01 $\mu$m in Type II--P (left panel), except for Si and FeS 
in the C--rich layer, and Al$_2$O$_3$, MgO and Fe$_3$O$_4$ in
the O--rich layer. The difference in the size of dust grains 
between Type II-P SN and Type IIb SN is 
much more clearly represented by 
the cumulative size distribution of dust by mass summed up over the
ejecta, which is given in Figure 5. Except for Al$_2$O$_3$, the dust
mass is dominated by grains with radii larger than $0.03 \mu$m in 
Type II--P SN (model
A; left panel), while the dust mass of all species  is dominated by the 
grains whose radii are less than 0.006 $\mu$m in Type 
IIb (model C; right panel). This 
leads to the big difference in the fate of dust grains in SNRs     
as is presented in the next section.  

\begin{table}[!t]
\caption{Dust mass formed in the the ejecta}
\smallskip
\begin{center}
{\small
\begin{tabular}{lccc}
\tableline
\noalign{\smallskip}
 model & A & B & C \\ 
\tableline
\noalign{\smallskip}
 Type & II-P & II-P & IIb \\  
 $M_{\rm pr}[M_{\odot}]$ & 15 & 20 & 18 \\ 
\tableline
\noalign{\smallskip}
dust species &  \multicolumn{3}{c}{$M_{{\rm d},j} [M_{\odot}]$ } \\ 
\tableline
\noalign{\smallskip}
C & $4.30\times 10^{-2}$ & $4.00\times 10^{-2}$ & $7.08 \times 10^{-2}$ \\ 
Al$_2$O$_3$ & $1.76 \times 10^{-3}$ & $4.33\times 10^{-2}$ & $6.19\times 10^{-5}$ \\
Mg$_2$SiO$_4$ & $2.95 \times 10^{-2}$ & 0.133 & $1.74 \times 10^{-2}$ \\
MgSiO$_3$ & $1.91\times 10^{-3}$ & $5.68 \times 10^{-3}$ & $5.46\times10^{-2}$ \\ 
MgO & $3.19 \times 10^{-7}$ & 0.159 & $2.36 \times 10^{-3}$ \\ 
SiO$_2$ & $2.06 \times 10^{-2}$ & $8.21 \times 10^{-2}$ & $1.57 \times 10^{-2}$ \\ 
Fe$_3$O$_4$ & $5.60\times 10^{-5}$ & $1.55 \times 10^{-3}$ & \null \\ 
FeS & $3.79\times 10^{-2}$ & $5.66 \times 10^{-2}$ & $1.47 \times 10^{-3}$ \\ 
Si & $6.65 \times 10^{-2}$ & $9.04\times 10^{-2}$ & $5.07 \times 10^{-3}$ \\
Fe & 0.134 & $6.46 \times 10^{-2}$ & \null \\
\tableline
\noalign{\smallskip}
total mass & 0.327 & 0.676 & 0.167 \\ 
\tableline
\noalign{\smallskip}
$M_{\rm d}/M_{\rm metal}$ & 0.695 & 0.202 & 0.129 \\ 
\tableline
\noalign{\smallskip}
\end{tabular}
}
\end{center}
\end{table}

\begin{figure}[!h]
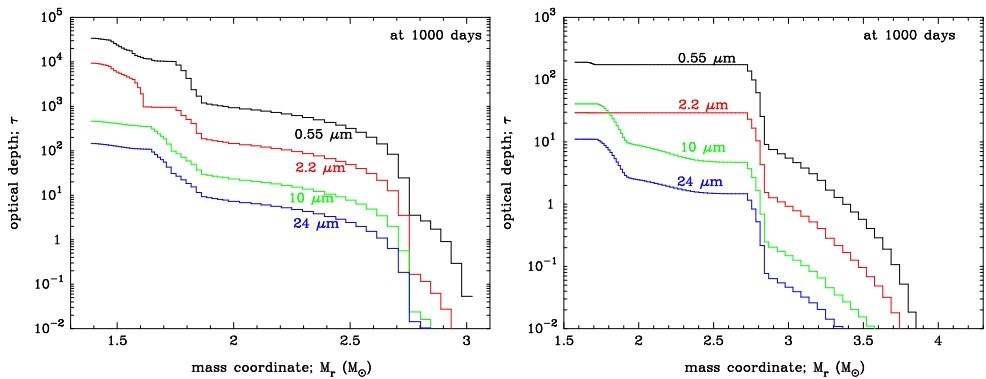

  \includegraphics[scale=.3, angle=270, totalheight=140pt, clip=true]
{figcolor/tau.15m.ps}
  \includegraphics[scale=.3, angle=270, totalheight=140pt, clip=true]
{figcolor/tau.93j.ps}
\caption{Optical depth of the ejecta at 1000 days after explosion
 as a function of mass coordinate; Left for Type II--P SN (model A) 
and right for Type IIb SN (model C).}
\end{figure}

Table 4 summarizes the mass of dust formed in the ejecta of
Type II--P SNe (model A and B) and Type IIb SN (model C). The masses of C,
Mg--silicates (Mg$_2$SiO$_4$ and MgSiO$_3$), SiO$_2$, Si and FeS 
do not depend so much  
on the model, while masses of Al$_2$O$_3$, MgO,
Fe$_3$O$_4$ and Fe are different from model to model, depending on the 
detailed elemental composition and density structure in the ejecta. 
Generally the total dust mass increases with increasing progenitor mass. 
The total dust mass $M_{\rm d}$ 
ranging from $\sim$ 0.17 to 0.68 $M_{\odot}$ is two
to three orders of magnitude larger than the mass inferred from the
observations.  The condensation efficiencies defined by $M_{\rm
d}/M_{\rm metal}$ ranges from 0.13 (model C) to 0.7 (model A). 
The condensation efficiency of model A is significantly higher than 
in the other models, because the metal mass 
within the He--core is 
smaller due 
to the thin O--rich layer (see Table 1 and Figure 1).    
Thus, being different from the total dust mass, the condensation
efficiency strongly depends on the detailed density structure 
and elemental composition in the 
ejecta.

Finally we present the optical depth of the ejecta at day 1000
after the explosion in Figure 6.
The optical depth of the ejecta after
dust formation decreases proportional  
to $t^{-2}$ since the ejecta
expands homologously. Thus, around day 500 after the onset of dust
formation, 
the ejecta is optically thick even at 24 $\mu$m  except for the outer
$\sim$ 0.3 $M_{\odot}$ (1.5 $M_{\odot}$) region for 
model A (model C). This 
result suggests that it could be inappropriate to estimate 
dust mass from the thermal emission by assuming the ejecta to 
be optically thin at MIR.
It takes
about 800 (60) yr at 0.5 $\mu$m for the entire ejecta to be 
optically thin, and about 50 (10) yr at
24 $\mu$m for model A (model C). 

\section{Evolution of dust in supernova remnant}

Dust grains formed in the ejecta suffer from erosion and
destruction by sputtering in the hot gas between the reverse and forward 
shocks produced by the interaction of
the ejecta with the surrounding medium. This process determines 
how much mass and what kind of 
dust can survive and be injected into 
ISM. In this section we demonstrate the
evolution of dust in SNRs, focusing on type IIb SN (model
C). Then, we calculate the time evolution of thermal emission from
dust in the SNR and compare with the observation of Cas A.

\subsection{Motion and destruction of dust in SNRs}

At the time when the ejecta encounters the ISM 
surrounding the SN, 
the forward shock propagating into the ISM and the reverse 
shock into the ejecta are generated (e.g. Truelove and McKee 1999), 
and the gas swept up by the shocks is heated up and ionized. The dust
grains hit by the reverse shock are 
injected into the shocked gas with a
velocity relative to the gas; gas is decelerated by 
the shock, while dust grains 
decoupling from gas penetrate into the hot gas with the same velocity
just before the encounter. Then dust grains are decelerated and
eroded by collision with gas. The deceleration of dust by the gas
drag is inversely proportional to $\rho_{\rm d}a$, and the erosion rate 
of dust by sputtering is almost independent of  
$a$ (e.g. 
Dwek \& Arendt 1992), where $a$ and $\rho_{\rm d}$ is 
the radius and the bulk density of dust, respectively. As a result, 
the small sized/low density dust is easily 
trapped into gas, and is 
eroded by non--kinetic (thermal) sputtering, while larger sized/high 
density dust moves through the gas, 
suffering kinetic sputtering. 

We calculate the motion and destruction of dust in the  
SNR, adapting 
the model for the SN explosion as 
the initial conditions for gas velocity and
density, and applying the results of dust
formation calculations as the initial spatial and 
size distributions of 
dust grains in the ejecta. The method of calculation and 
the underlying
assumptions are the same as those by Nozawa et al. (2007). 
In the calculations, we assume that the
ejecta expands into a uniform medium with solar metallicity. 
We consider the two cases for the ambient
hydrogen number density; $n_{\rm H}$ =1.0 and 10.0 cm $^{-3}$. We assume 
the gas temperature in the ambient medium $T_{\rm g,amb}$ = $10^4$ K, and that 
the ejecta hits the ISM at $t_{\rm enc}=10$ yr after the explosion. Note that 
the assumptions on $T_{\rm g,amb}$ and $t_{\rm enc}$
do not significantly affect the result of
calculations. The sputtering yield of each grain species is taken from
Nozawa 
et al. (2006).

\begin{figure}[!ht]
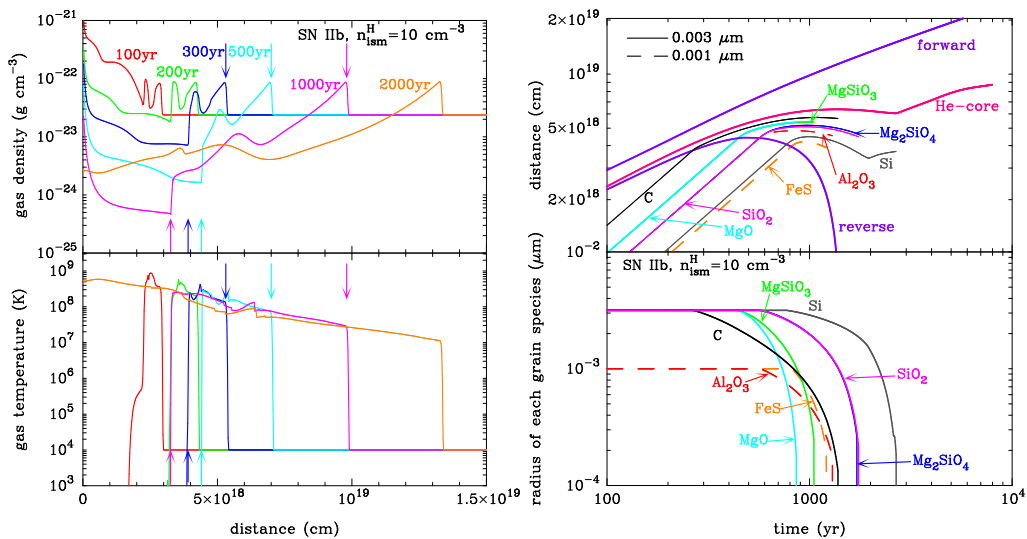

  \includegraphics[scale=.3, angle=0, totalheight=200pt, clip=true]
  {figcolor/93j-1c10n-rho-t.ps}
  \includegraphics[scale=.3, angle=0, totalheight=200pt, clip=true]
{figcolor/93j-1c10n-r-a.ps}
\caption{Time evolution of gas density and temperature (left panel), and
 the motion and destruction of dust grains (right panel) in the Type IIb
 SNR for $n_{\rm H}= 10.0$ cm$^{-3}$ }
\end{figure}

The left panel of Figure 7 shows the time evolution of gas density and 
temperature in the SNR of type IIb SN (Type IIb SNR) for 
$n_{\rm H}=$ 10 cm$^{-3}$; 
the positions of the 
reverse (forward) shock at given times are  
indicated by upward (downward) arrows.
The temperature of gas hit by the shocks quickly rises   
up to 10$^{7}$ to 10$^9$ K, and is kept to be $ > 10^5$ K until the cool dense
shell forms behind the forward shock at $t \sim 10^6$ yr after the
explosion. 
The reverse shock encounters  
the C--rich layer at $\sim 80$ yr,   
the O--rich layer at $\sim 440$ yr,  
the Si--S--Fe rich layer at $\sim 800$ yr,  
and the Fe-Si--S rich layer at $\sim 880$ yr after the explosion, and 
the  ejecta are completely swept up by 
the reverse shock in $\sim 1400$ yr. The right--upper panel
shows the trajectories of dust species with radii of 0.001 $\mu$m
(dashed curves) and 0.003 $\mu$m (solid curves) together with the 
positions of the forward and reverse
shocks and the outer boundary of He--core.   
The time evolution of the radius of each grain 
species is depicted in the 
right--lower panel. As can be seen, grains with radii of 
0.003 $\mu$m are quickly trapped into the hot gas behind the reverse 
shocks and are completely destroyed by thermal sputtering within $\sim 3 
\times 10^3$ yr. How long dust grains remain in SNR depends 
not only on the initial position but also on the chemical composition.  

The time evolution of the mass of each dust species 
in the hot gas of Type IIb SNR is displayed in 
Figure 8; Left for $n_{\rm H}=1.0$ cm$^{-3}$ , and right for 
$n_{\rm H}=10.0$ cm$^{-3}$. Except for C grains for $n_{\rm
H}=1.0$ cm$^{-3}$, all dust grains are destroyed in  
$ t< 3 \times 10^4$ ($4 \times 10^3$) yr for 
$n_{\rm H}=1.0 ~(10.0)$ cm$^{-3}$. 
Although the time necessary for the 
reverse shock to reach the dust forming layer is almost two times longer for 
$n_{\rm H}=1.0$ cm$^{-3}$ than for $n_{\rm H}=10.0$ cm$^{-3}$,   
it takes much longer for dust
grains to be destroyed for $n_{\rm H}=1.0$ cm$^{-3}$ because 
the lower density in the hot gas caused by the delayed encounter 
reduces the erosion rate of dust by sputtering significantly. 
It should
be noted that C grains dominate the dust mass almost 
over all times for $n_{\rm H}=1.0$ cm$^{-3}$, while for
$n_{\rm H}=10.0$ cm$^{-3}$ the mass of  
C grains quickly decreases  
at $t>300$ yr, 
and then MgSiO$_3$ and SiO$_2$ grains dominate the dust mass in 
the SNR. Thus, 
the ambient gas density strongly affects  
not only the mass but also the 
composition of dust grains remaining in the SNR. 

\begin{figure}[!t]
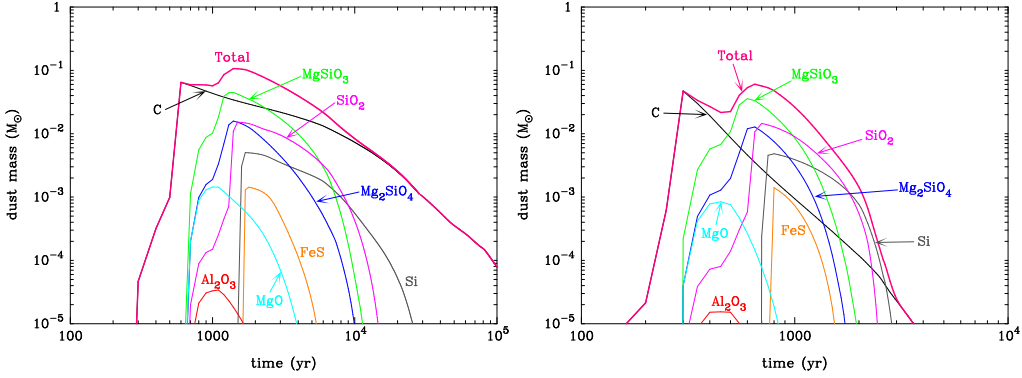

  \includegraphics[scale=.4, angle=270, totalheight=140pt, clip=true]
{figcolor/93jn1.mass.ps}
  \includegraphics[scale=.4, angle=270, totalheight=140pt, clip=true]
{figcolor/93jn10.mass4.ps}
\caption{Time evolution of dust mass in the shocked gas of 
the Type IIb SNR:
Left for $n_{\rm
 H}=1.0$ cm$^{-3}$ and right for $n_{\rm H}=10.0$ cm$^{-3}$.}
\end{figure}

\begin{table}
\caption{Mass of dust surviving in the SNRs $M_{\rm
 surv}$ at t=10 
$\time
 10^5$ yr and the efficiency of survival 
defined by $\eta=M_{\rm surv}/M_{\rm d}$}
\smallskip
\begin{center}
{\small
\begin{tabular}{lcccc}
\tableline
\noalign{\smallskip}
$n_{\rm H}[{\rm cm}^{-3}]$ & \multicolumn{2}{c}{1.0} & 
\multicolumn{2}{c}{10.0} \\
\noalign{\smallskip}
\tableline
\noalign{\smallskip}
 model (type) & $M_{\rm surv} [M_{\odot}]$ & $\eta$ &  $M_{\rm surv}
 [M_{\odot}]$ & $\eta$ \\
\noalign{\smallskip}
\tableline
\noalign{\smallskip}
 A (Type II--P) &  0.190 & 0.581 & 0.104  & 0.318 \\
 B (Type II--P)  &  $5.66 \times 10^{-2}$  & $8.37 \times 10^{-2}$ & $3.91 
\times 10^{-2}$  & $5.78 \times 10^{-2}$ \\
 C (Type IIb \.)   & $7.83 \times 10^{-5}$ & 6.07 $\times 10^{-4}$ & 0.00
 & 0.00 \\ 
\noalign{\smallskip}
\tableline
\end{tabular}
}
\end{center}
\end{table}

Table 5  
presents the surviving dust mass $M_{\rm surv}$ at $t=10^5$ yr for 
the models. Although $M_{\rm surv}$ decreases with increasing
$n_{\rm H}$, how much mass of dust survives strongly 
depends on the type of SNe through the thickness of outer H--envelope 
that influences    
the size of dust formed in the ejecta as well as 
the time when the reverse shock reaches the dust forming layers. 
In the Type IIb SNR, dust is destroyed quickly partly because 
the small sized dust grains with radii $<$ 0.006 $\mu$m populate 
the remnant region. In addition, 
the earlier arrival of reverse shock at   
the dust--forming layer causes  
the gas density behind the reverse shock 
to be higher, and  enhances the erosion rate of dust grains; 
for example, the 35
times earlier arrival to the C--rich layer compared to the Type II--P 
SNR (model A) results in the 10 times higher gas 
density behind the reverse shock in the Type IIb SNR. 
However, it should be noted that, contrary to the expectation, 
dust grains are more destroyed in model B than in model A, 
despite that   
the mass of H--envelope is almost same. 
One of the reasons is that MgO grains with a 
sputtering rate somewhat larger than others 
grains (Nozawa et al. 2006) 
are more abundant in model B (see Table 4), 
and another is that the sizes of Fe and Si 
grains in model A are  
significantly larger than those in model B. Thus, the survival of dust
grains in SNR depends on not only the thickness of H--envelopes but
also on the detailed chemical composition, size
distribution, and the
amount of dust grains formed in the ejecta.

The efficiency of
survival $\eta$ defined as $M_{\rm surv}/M_{\rm d}$ is larger 
than $ \sim 0.06$ in the Type II--P SNR, but less 
than 6 $\times 10^{-4}$ in the Type IIb SNR. 
Dust grains surviving at $t=10^5$ yr
are almost free from the destruction 
by sputtering, and are trapped into 
the dense cool shell formed behind 
the forward shock or
injected into the ISM. Thus, the injection efficiency 
strongly depends on the type, which  
should be incorporated into 
chemical evolution models of dust in the ISM.  
In addition, it should be
mentioned here that the survival in SNRs and the 
injection into the ISM of newly formed dust
depend on the density structure in the ISM as 
well as the inhomogeneity in 
the ejecta. For the ambient medium whose gas density decreases as $r^{-n}$ 
with distance $r$, the time evolution of the density of gas swept up 
by the reverse shock is different from that for a constant density as shown
by Chevalier \& Oishi (2003), which affects the evolution of dust in 
SNRs. If dust grains reside in dense clumps (knots) as suggested by the 
observations (e.g. Arendt, Dwek, \& Moseley 1999), 
the dust grains could survive, 
since the reverse shock encountering   
dense clumps could indeed be radiative so that 
the gas temperature could stay cool enough 
for dust grains to be rescued from the  
destruction  by 
sputtering. These aspects should be explored to reveal 
how much mass of dust grain can survive in SNRs  
and be injected into the ISM. 

\subsection{Thermal emission from dust in Type IIb SNR}

Dust grains formed in the ejecta and injected
 into the hot plasma between the reverse and 
forward shocks are heated up by collision with gas, and emit 
thermal   
radiation. The equilibrium temperature of dust grain is determined by 
balancing the collisional heating with the radiative cooling. However, 
in a rarefied hot plasma such as in SNRs, 
small sized dust grains undergo stochastic heating 
which  
affects the emissivity and the resulting spectral energy
distribution (SED); see Dwek (1986) and Dwek et al. (2008) for details. 

\begin{figure}[!hb]
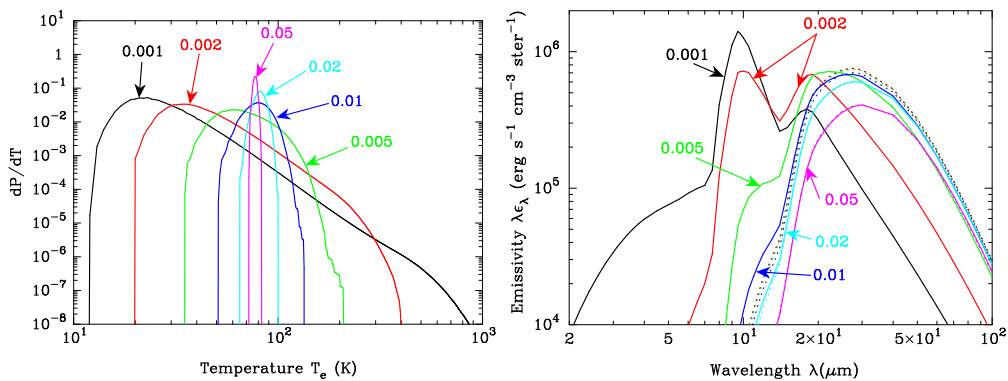

  \includegraphics[scale=.4, angle=270, totalheight=140pt, clip=true]
{figcolor/ptdist.pr.ps}
  \includegraphics[scale=.4, angle=270, totalheight=140pt, clip=true]
{figcolor/emiss.pr.ps}
\caption{Temperature distribution functions of small astronomical
 silicate in 
the hot plasma with $n_{\rm e}=10 {\rm cm}^{-3}$ and $T_{\rm e}=10^7$ K (left) and the
emissivities (right) for given radii, where the radius is in units of $\mu$m.}
\end{figure}   

For example, Figure 9 shows the
temperature distribution function and the emissivity of astronomical 
silicate (Draine and Lee 1984) embedded in a hot plasma with the
electron 
temperature of $T_{\rm e}$ = $10^7$ K and 
number density of $n_{\rm e}$ 
= $10.0$ cm$^{-3}$, which is calculated by a
Monte--Carlo method. The fraction of incident 
energy that is deposited in a grain is calculated by fitting 
the experimental data of the stopping ranges of electrons tabulated by 
Iskef et al. (1983), according to the method by Dwek (1987b).
With decreasing radius, 
the temperature distribution function 
becomes broader and the 
resulting emissivity deviates strongly from 
the emissivity calculated by
the equilibrium temperature; the deviation becomes significant for the
radius of $a < 0.02 \mu$m.

Taking into account the stochastic heating,  
we demonstrate the time evolution of thermal emission from dust
grains embedded in the shocked gas in the Type 
IIb SNR, based on 
the results of dust formation and evolution calculations for Type IIb SN
presented in \S ~3 
and \S ~4.1.  In the calculations, the heat capacities of grain species 
are taken from Takeuchi et
al. (2005), and we use the same optical constants as those of 
Hirashita et al. (2005), except for amorphous 
Al$_2$O$_3$ (Begemann et al. 1997 for the wavelength 
$\lambda > 8 ~\mu$m), SiO$_2$ (Phillip 1985) and Si (Pillar 1985). 
We consider only collision with electrons for heating, because 
the radii of dust grains formed in the ejecta are so small that dust
grains injected into the hot gas are quickly trapped 
and comove with the 
gas, and the heating is dominated by collision with electrons. In order
to compare with the observations of Cas A, we put the SNR at the
distance of 3.4 kpc.      
Figure 10 and 11 show the time evolution of SED 
ranging from 200 to 1800 
yr after the explosion for $n_{\rm H}$=1.0 and 10.0 cm$^{-3}$, 
respectively, where the solid (dotted) curve denotes the SED with (without) 
stochastic heating. The solid circles are the flux densities
subtracting the synchrotron radiation from the observed ones  
tabulated by Hines et al. (2004).  

The SEDs taking into account the
stochastic heating is  
completely different from those with the
equilibrium temperatures, especially in the wavelengths shorter than 20
$\mu$m. After the revere shock encounters   
the C--rich layer
at $\sim$ 180 (80) yr for $n_{\rm H}$=1.0 (10.0) cm$^{-3}$, the thermal 
emission from the stochastically heated C grains contribute to 
the SED until  
the reverse shock encounters the O--rich layer. 
In the case of $n_{\rm
H}=1.0$ cm$^{-3}$, the thermal emission from C grains 
dominates the SED over the entire simulation period considered 
in the calculation, 
because the mass of dust in the SNR is almost 
dominated by C grains as 
can be seen from Figure
8. The emission features of MgO 
appearing at $\lambda \sim$ 16 $\mu$m from  $t=800$ to 1200 yr,   
Mg--silicates at $\lambda \sim$ 10 $\mu$m 
after $t=1200$ yr and SiO$_2$ at $\lambda \sim$ 20 $\mu$m 
after $t=1600$ yr are not prominent, being overwhelmed by thermal 
emission from C grains.  
On the other hand, in the case of $n_{\rm H}=10.0$ cm$^{-3}$, 
with C grains being more efficiently destroyed 
by sputtering, the
emission features from other dust grains get 
prominent with time, 
after the reverse shock encounters the O--rich 
layer at $t \sim 400$
yr; although the
feature of MgO is weak, the emission feature of
Mg--silicates around 10 $\mu$m becomes noticeable at
$t=600$ yr, after then the 20 $\mu$m feature of SiO$_2$ is prominent at 
$t=800$ yr. With increasing time, the SED is dominated by the
emission features from stochastically heated dust grains.

\begin{figure}
\plotone{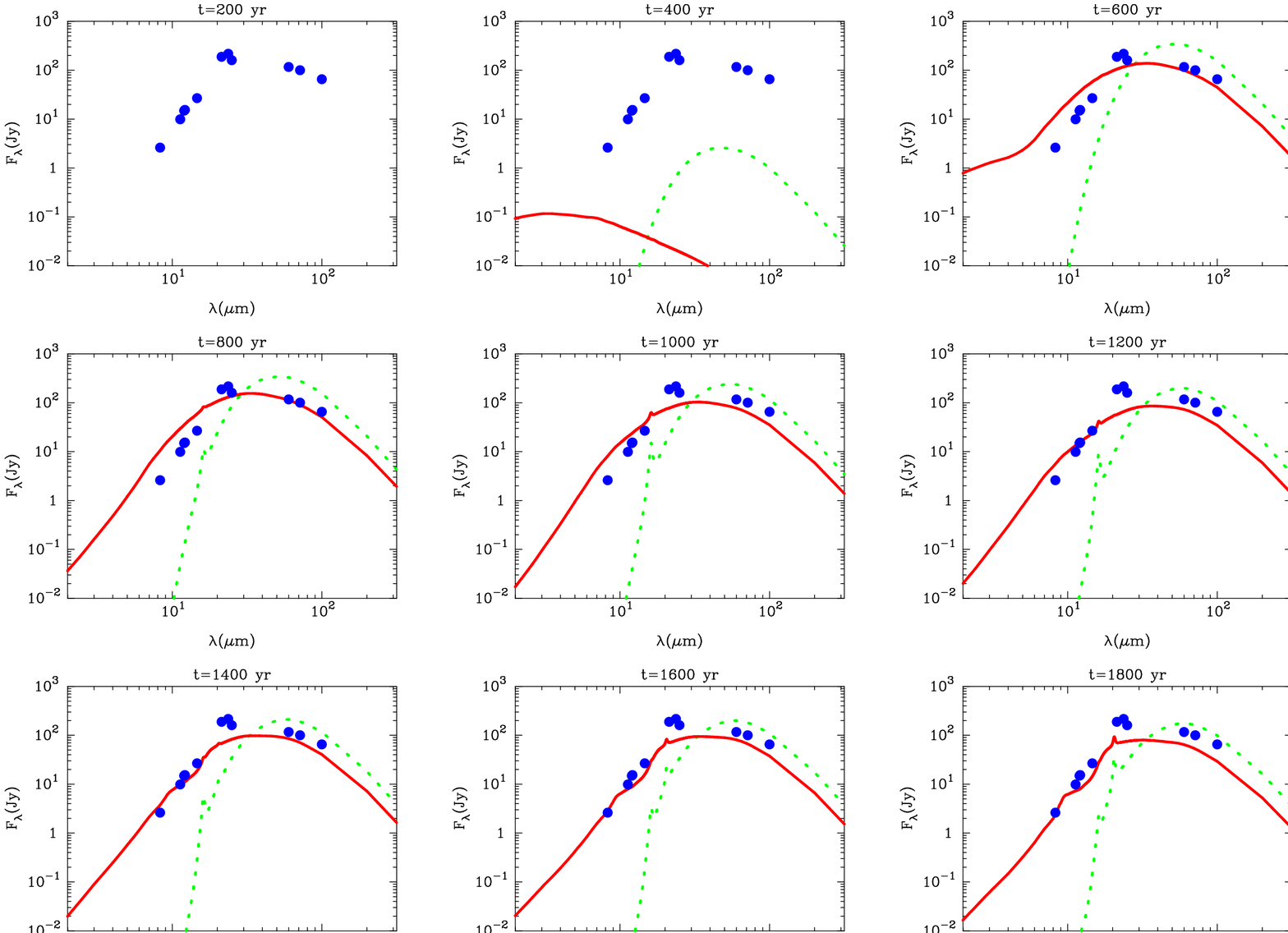}
\vspace{0.8cm}
\caption{The time evolution of SEDs of Type IIb SNR for 
$n_{\rm H}=1.0$ cm$^{-3}$ with (solid curve) and without (dotted
 curves) stochastic heating. Closed circles are the flux
 densities subtracting the synchrotron radiation 
from the observed ones of Cas A (Hines et al. 2004)} 
\vspace{0.5cm}
\plotone{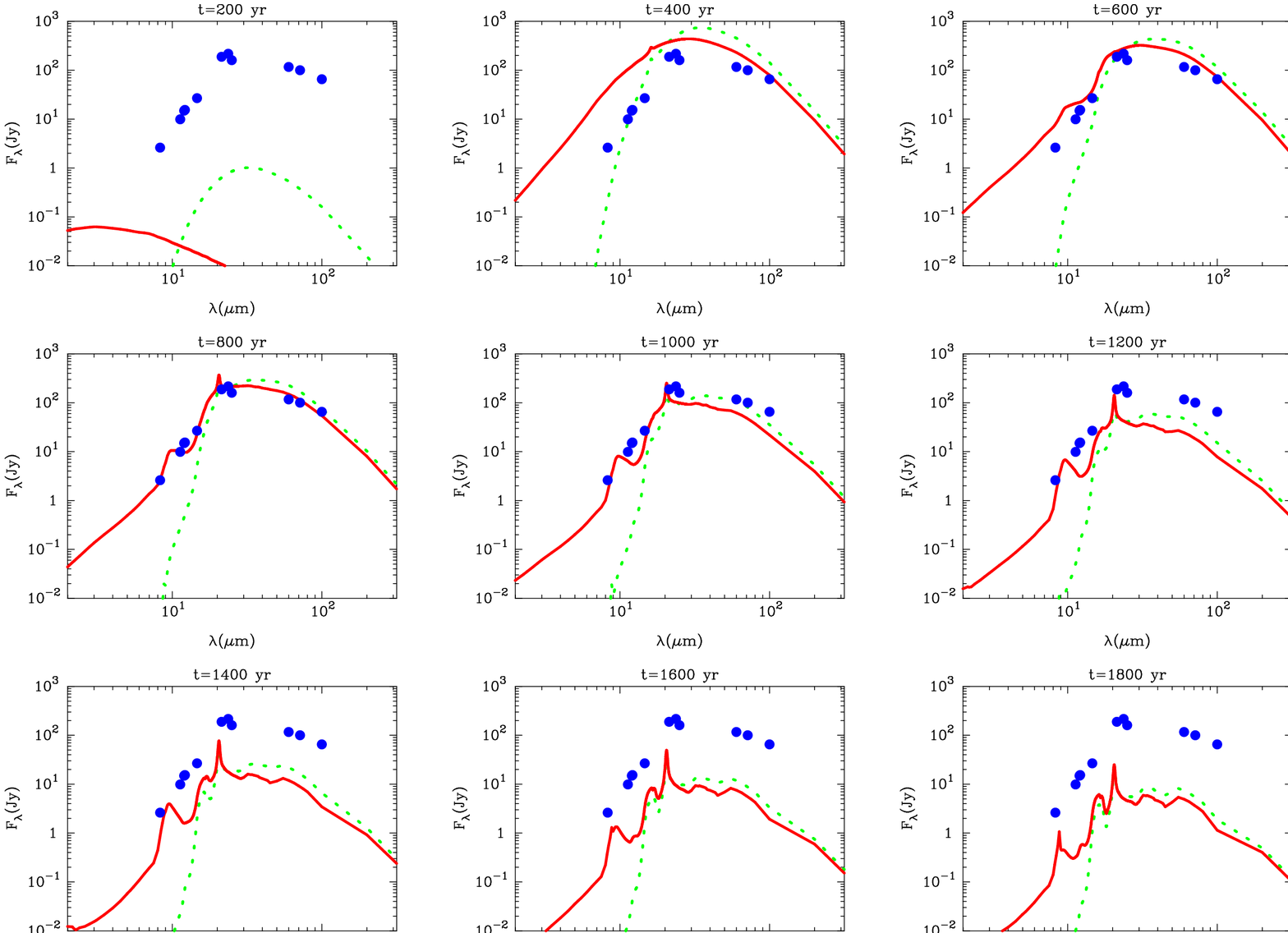}
\vspace{0.8cm}
\caption{The same as Figure 10, but for $n_{\rm H}=10.0$ cm$^{-3}$} 
\end{figure}

The result of calculations clearly demonstrates that the time evolution
of the SED is very sensitive to the evolution of 
dust in SNRs through 
the destruction by sputtering and the stochastic heating, 
which strongly depends on the density in the ambient medium.  
It should be noted that
the calculated SED cannot be reproduced by a single dust component 
with multiple temperatures.

\subsection{Comparison with the observation of Cas A}

\begin{figure}[!h]
\begin{center}
  \includegraphics [scale=0.3, angle=270, totalheight=240pt, clip=true]
{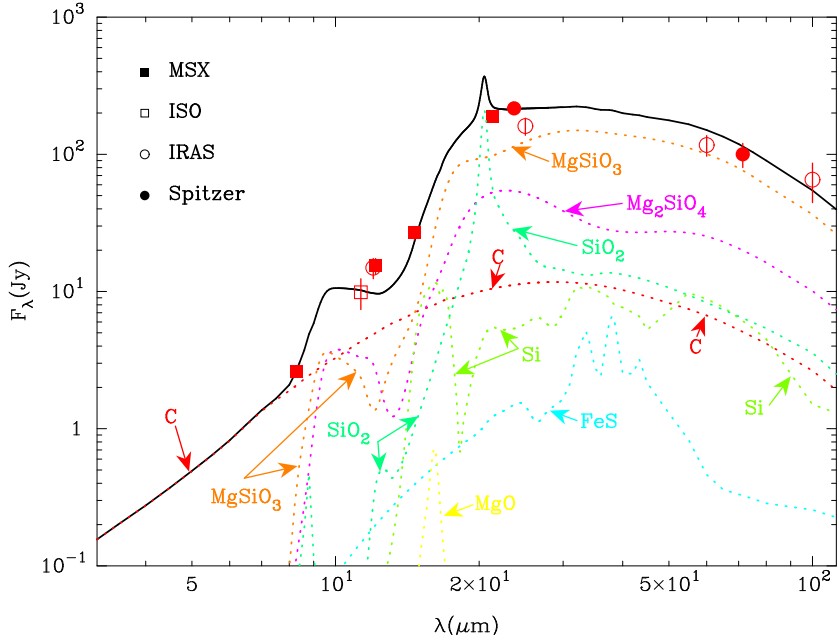}
\caption{Comparison with the observation of Ca A. The symbols denote 
the flux
 densities of Cas A. The solid curve is the
 total flux density, and the dotted curves denote 
 the contributions of each grain species.}
\end{center}
\vspace{-0.5cm}
\end{figure}

Cas A is the nearby Galactic SNR well--studied at various wavelengths 
and was recently identified as Type IIb SN 
from the spectrum of the scattered 
light echo (Krause et al. 2008). The presence of dust formed in
the ejecta  
has been confirmed by the infrared and submillimeter 
observations (Lagage et al. 1996; Arendt et al. 1999; Dunne et al. 2003). 
However, the dust mass derived from the observations has
been controversial, ranging from $< 1 \times 10^{-3}$ to $\sim 1.0 M_
{\odot}$ (Dwek et al. 1987a; Douvion, Lagage, \& Pantin 2001; 
Dunne et al. 2003; 
Hines et al. 2004; Rho et al. 2008). The comparison of the calculated
SED with the observations of Cas A could provide some insights on dust
and its evolution in the Cas A SNR, 
although the models of Type IIb 
SN and the evolution of the SNR  
are not necessarily applicable to the Cas A SNR.

As shown in Figures 10 and 11, the calculated SEDs for 
$n_{\rm}=1.0$ cm$^{-3}$, being dominated by 
thermal emission from small C grains, cannot reproduce the 
observations at the wavelengths $\lambda <$ 20 $\mu$m at $t \le 1000$
yr, and are significantly smaller than the observed flux densities  
at $\lambda \ge$ 20 $\mu$m. On the other hand, 
for $n_{\rm H}=10.0$ cm$^{-3}$, the calculated SED at $t=800$ 
around which the reverse shock encounters  
the Si--rich layer seems to reproduce the observed SED.

Figure 12 shows the detailed comparison of the SED calculated 
at $t=$ 800 yr with the observation and 
the contribution of each dust grains. Although the 
discrepancy around $\lambda \sim$ 12 $\mu$m is remarkable and the 20 
$\mu$m feature of SiO$_2$ is too sharp, the calculated SED 
can reasonably reproduce the overall  
shape of the observed SED 
without any tuning. The SED is dominated by Mg-silicates 
and SiO$_2$ at $\lambda >$ 15 $\mu$m, and
by C and Mg-silicates at $\lambda <$ 10 $\mu$m. The dominant dust species
(the mass in unit of solar mass) are MgSiO$_3$ ($1.72 \times 10^{-2}$), 
SiO$_2$ ($1.08 \times 10^{-2}$), MgSiO$_4$ ($6.46 \times 10^{-3}$), 
Si ($3.62 \times 10^{-3}$), C ($1.6 \times 10^{-3}$), and 
FeS ($1.04 \times 10^{-3}$). The total dust mass is $\sim$ 
0.04 $M_{\odot}$, which is consistent with the dust mass 
of 0.02--0.054 $M_{\odot}$ evaluated by Rho et al. (2008),  
apart from the
details. The presence of Si grains in the model 
indicates that the reverse shock already reaches  
a part of the 
inner Si--S--Fe layer, which is also not 
in contradiction with the observation
of Cas A by Spitzer showing that the deeper Si--rich 
layer has been hit by the reverse shock 
in the region  associated with
the jet arising from the asymmetric explosion (Ennis et al. 2006).  

\section{Summary and concluding remarks}

We calculate dust formation in the ejecta and evolution in the SNRs of 
Type II--P and IIb SNe in order to clarify how 
these precesses depend
on the type of CCSNe through the difference in the
thickness (mass) of the outer envelope. We show that the mass of newly 
formed dust ranging from $\sim 0.1$ to 0.7 $M_{\odot}$ does not so
much depend on, but the size of dust grains is sensitive to 
the thickness of H--envelope. The radii of dust grains are 
less than 0.006 $\mu$m 
in Type IIb SN with
$M_{\rm H-env}=$ 0.08 $M_{\odot}$, while the dust mass is dominated by the 
grains with radii larger than 0.03 $\mu$m in Type II--P SNe with 
$M_{\rm H-env}
\sim 10 M_{\odot}$. The difference in the size of the 
dust formed in the 
ejecta plays a crucial role in 
the evolution of dust in SNRs because the 
smaller sized grains are quickly trapped into the hot gas behind the 
reverse shock and 
are destroyed by sputtering. The surviving dust mass ranges 
from  0.19 to $3.9\times 10^{-2}
M_{\odot}$ in the SNR of Type II--P SN for the ambient H number density 
of $n_{\rm H}$ = 1.0--10.0  
cm$^{-3}$, but the dust grains are almost 
completely destroyed in the SNR of Type IIb SN.  

The SED of the thermal emission from dust embedded in SNRs   
can be applied as a diagnosis of  
the evolution of dust, reflecting the destruction by sputtering and 
the stochastic heating. The comparison of the calculated SED of Type IIb
SNRs with the
observation of Cas A suggests that the mass of dust in the SNR is 0.04
$M_{\odot}$, which implies that the mass of dust formed in the ejecta is 
not less than
$1.0 \times 10^{-3} M_{\odot}$, being different from the dust mass
estimated from the dust forming CCSNe. 
In addition, Mg--silicates and SiO$_2$ with mass reaching up to  
$3.0 \times 10^{-2} M_{\odot}$ 
dominate the SED 
at $\lambda >15 \mu$m, in contrast to the observed 
dust forming SNe that so far show no signature of silicates. 
This may indicate that these dust grains formed in the O--rich layer cool 
down quickly 
and evade being detected by the observations as is discussed 
by Nozawa et al. (2008). 

It has been claimed that the mass of dust formed per  CCSN is less   
than $1.0 \times 10^{-3} M_{\odot}$ and is too small to 
contribute to the
inventory of dust in ISM, based on the observations. 
However,  it should bear in mind that the 
conclusion comes from a small number of observations. In order to
clarify when how much mass and what kind of dust 
condenses in the ejecta
of various types of SNe, it is inevitable to monitor SNe with temporal and
wavelength coverages as wide as possible. In addition, the dedicated
radiation transfer calculations such as those 
by Sugerman et al. (2006)
and Ercolano et al (2007) are  
promising in evaluating the dust mass
compared with the observations, taking into account 
the distribution and chemical composition of 
dust grains in the ejecta.

\acknowledgements 

This work is partly supported by the Grant--in--Aid for Scientific
Research of the Japan Society for the Promotion of Science 
(18104003, 20340038).


\end{document}